\documentclass[nofootinbib,showpacs,aps,prc,reprint,floatfix]{revtex4-1}
        \usepackage{color}
\usepackage{epsfig}
\usepackage{graphicx}
\usepackage{amssymb}

\begin{document}

\title{Aspects of the momentum dependence of the equation of state
       and of the residual $NN$ cross section, and their effects on
       nuclear stopping}
\author{Z.\ Basrak$^1$}
\email{basrak@irb.hr}
\author{P.\ Eudes$^2$}
\email{eudes@subatech.in2p3.fr}
\author{V.\ de la Mota$^2$}
\affiliation{$^1$Ru{d\llap{\raise 1.22ex\hbox
   {\vrule height 0.09ex width 0.2em}}\rlap{\raise 1.22ex\hbox
   {\vrule height 0.09ex width 0.06em}}}er
   Bo\v{s}kovi\'{c} Institute, Zagreb, Croatia}
\affiliation{$^2$SUBATECH, EMN-IN2P3/CNRS-Universit\'e de Nantes,
          Nantes, France}

\date{\today}

\begin{abstract}
With the semiclassical Landau-Vlasov transport model we studied
the stopping observable $R_E$, the energy-based isotropy
ratio, for the $^{129}$Xe\,+\,$^{120}$Sn reaction at beam
energies spanning 12$A$ to 100$A$ MeV.
We investigated the impacts of the nonlocality of the nuclear
mean field, of the in-medium modified nucleon-nucleon ($NN$)
cross section and of the reaction centrality.
A fixed set of model parameters yields $R_E$ values that
favorably compare with the experimental ones, but only for
energies below the Fermi energy $E_F$.
Above $E_F$ agreement is readily possible, but by a smooth
evolution with energy of the parameter that controls the
in-medium modification of $NN$ cross section.
By comparing the simulation correction factor ${\cal F}$
applied to the free $NN$ cross section with the one deduced
from experimental data
[Phys.\ Rev.\ C\,{\bf 90}, 064602 (2014)], we infer that the
zero-range mean field almost entirely reproduces it.
Also, in accordance with what has been deduced from
experimental data, around $E_F$ a strong
reduction of the free $NN$ cross section is found.
In order to test the impact of sampling central collisions
by multiplicity an event generator (HIPSE) was used.
We obtain that high multiplicity events are spread over a
broad impact parameter range, but it turns out that this has a
small effect on the observable $R_E$ and, thus, on ${\cal F}$
as well.
\end{abstract}

\maketitle

\section{\label{sec1}Introduction}
The ratio between transverse and longitudinal components of
kinematical observables is a measure of the conversion of
the initial entrance channel motion into intrinsic degrees
of freedom in heavy-ion collisions (HICs).
Such an observable gives an insight on the rate of a system's
equilibration, of the dissipation of the available energy,
as well as of HIC stopping power\ \cite{stroe83,bauer88}.
Thanks to such an observable, the FOPI Collaboration
has evidenced partial nuclear transparency in HIC in the
beam-energy range $E_{\rm inc}\!\approx$\,0.1$A$\,--\,1$A$
GeV\ \cite{fopi}.
More recently,
by examining the ratio of transverse to longitudinal energy
$R_E$ and linear momentum $R_p$ for the most violent HICs,
the INDRA Collaboration has
revealed a substantial reduction of the nuclear stopping
power at $E_{\rm inc}\!\approx$\,10$A$\,--\,100$A$ MeV\
\cite{lehaut}.
This stopping observable reaches a minimum around the Fermi
energy $E_F$ and stagnates or very weakly increases with the
further increase of $E_{\rm inc}$ at least up to 100$A$ MeV,
the upper limit of the energy range available in this study.
The above observation is valid for all (mass symmetric)
systems studied, with system masses
$A_{\rm sys}$\,=\,72\,--\,394 a.m.u.
It is worth emphasizing that the fusion cross section normalized
by the total reaction cross section exhibits an analogously
rapid fall-off up to about $E_F$\ \cite{eud13,eud14}, a
behavior especially evident for mass-symmetric systems (cf.\
Fig.\ 6 of Ref.\ \cite{eud14}).

In a recent publication\ \cite{lopez14} the above observable
$R_E$ was analyzed for the $Z$\,=\,1 subset of the same
INDRA data.
The $Z$\,=\,1 $R_E$ displays a slightly stronger increase
with $E_{\rm inc}\!>\!E_F$ for the heavier systems\
\cite{lopez14} relative to the $R_E$ values obtained in the
previous study\ \cite{lehaut} which included light
charged particles and fragments, but also was somewhat
more stringent on the selection of the most central events.
The authors of Ref.\ \cite{lopez14} report a minimum of $R_E$
around $E_F$, which is particularly enhanced when $R_E$ is
normalized to the Fermi-gas-model prediction of the incoming
$R_E$ value at a given $E_{\rm inc}$.\ \cite{su13}

In Ref.\ \cite{lopez14} it was assumed that protons are
predominantly dynamically emitted during the early reaction
phase, in accordance with Refs.\ \cite{eud97,had99}.
Such a hypothesis offers a possibility of extracting
information on the in-medium correction for the free
nucleon-nucleon ($NN$) cross section
$\sigma^{\rm free}_{\rm NN}$.
Following such an argument, starting from the
experimental $R_E$ values in Ref.\ \cite{lopez14},
with some basic assumptions about the effects owing to the
Pauli-exclusion principle, the nucleon mean free path was
extracted and an effective value of the in-medium $NN$ cross
section $\sigma^m_{\rm NN}$ was deduced.
In the process, a correction factor ${\cal F}$ was obtained
by which $\sigma^{\rm free}_{\rm NN}$ has to be multiplied at
each $E_{\rm inc}$ to get a proper $\sigma^m_{\rm NN}$ value.
The authors found that (i) a \textit{significant reduction}
of $\sigma^{\rm free}_{\rm NN}$ is present in HICs below 100$A$
MeV and (ii) this change of $\sigma^{\rm free}_{\rm NN}$ is
strongly dependent upon $E_{\rm inc}$.
At the lowest energies the measured $R_E$ is compatible with
the full stopping value ($R_E\!\approx$\,1) and the effective
$\sigma^m_{\rm NN}$ amounts to about
0.4\,$\sigma^{\rm free}_{\rm NN}$.
One should keep in mind that the authors claimed a large
uncertainty on the factor ${\cal F}$ below
$E_{\rm inc}\!\sim$\,30$A$ MeV, a subject for which they have
announced a devoted publication\ \cite{lopez14}.
At incident energies around $E_F$ where $R_E$ attains its
minimum, $\sigma^m_{\rm NN}$ is reduced to less than one fifth
the free $\sigma^{\rm free}_{\rm NN}$ value (${\cal F}$\,=\,0.17)
and then the effective $\sigma^m_{\rm NN}$ steadily and regularly
increases up to half of the free value (${\cal F}\!\approx$\,0.5)
at $E_{\rm inc}$\,=\,100$A$ MeV\ \cite{lopez14} (see also
Fig.\ \ref{cf} in the present work).

The stopping observables $R_E$ and/or $R_p$ have also been
investigated in isospin-dependent quantum molecular dynamics
(IQMD)\ \cite{su13,liu01,zhang,kaur11,vin12} and antisymmetrized
molecular dynamics (AMD)\ \cite{zhao} model studies of HICs.
All these works were carried out before publication of
Ref.\ \cite{lopez14}.
Neither of simulation approaches predicts the remarkable
in-medium reduction of $\sigma^{\rm free}_{\rm NN}$ found in
Ref.\ \cite{lopez14}.
In the AMD study, specific attention has been paid to performing
the analysis by meticulously following the experimental
procedure for data handling\ \cite{zhao}.
The simulation with in-medium $\sigma_{\rm NN}$ due to
Li and Machleidt\ \cite{li93-4} (the free $\sigma_{\rm NN}$)
undershoots (overshoots) the data of\ \cite{lehaut}.
An agreement with the data could only be reached at
$E_{\rm inc}\!\ge$\,80$A$ MeV by doubling the theoretically
established $\sigma^m_{\rm NN}$ of\ \cite{li93-4}.
A systematic investigation of the impact of $\sigma_{\rm NN}$
on $R_E$, however, has not been performed yet.
The intention of the present study is twofold:

\noindent
1. by varying the nuclear equation of state (EOS) and the
parametrization of $\sigma_{\rm NN}$, to investigate how
the semiclassical Landau-Vlasov (LV) transport model of HIC\
\cite{gre87,seb89} complies with the experimentally deduced
dependence of the stopping observable $R_E$ on $E_{\rm inc}$
and

\noindent
2. by varying a simple multiplicative factor ${\cal F}$ of
the free $NN$ cross section, to compare thus obtained values
for ${\cal F}$ with those reported in Ref.\ \cite{lopez14}.

\section{\label{sec2}Model ingredients}
Within the semiquantal extension of the Boltzmann transport
theory, the highly nonlinear LV equation governs the
spatio-temporal evolution of the one-body density
distribution function $f({\bf r},{\bf p};t)$:
\begin{equation}
{\frac{\partial f({\bf r},{\bf p};t)}{\partial t}} +
\{ f({\bf r},{\bf p};t),H \}=I_{\rm{coll}}(f({\bf r},{\bf p};t)) ,
\label{LVeq}
\end{equation}
\noindent
which gives the probability of finding at the instant $t$ a
particle in the phase-space point $({\bf r},{\bf p})$.
$\{\;,\;\}$ stands for the Poisson bracket, whereas
$H$ is the one-body Hamiltonian describing the Coulomb
potential and the nuclear mean field.
We present the results obtained with a soft nonlocal mean field
labeled D1-G1 ($K_\infty$\,=\,228 MeV, $m^\ast/m$\,=\,0.67) due
to Gogny\ \cite{dech80} and those obtained with the standard
simplification of the soft zero-range Skyrme interaction due
to Zamick ($K_\infty$\,=\,200 MeV, $m^\ast/m$\,=\,1.0)\
\cite{zam73}.
The D1-G1 force is reputed to reproduce fundamental
properties of nuclear matter as well as those of finite
nuclei\ \cite{dech80} while the Zamick parametrization of
the EOS is, owing to its simplicity, of rather widespread use in
a number of microscopic approaches.
Details on both the nonlocal and the local parametrizations
of the used EOS may be found in Tables I and III of Ref.\
\cite{had95}, respectively.
We have demonstrated that the LV model is able to correctly
describe experimental observations in the intermediate energy
regime\ \cite{eud97,had99,seb89,had95,lv_g}.
The use of only a density dependent EOS is
legitimated by the finding\ \cite{liu01,vin12} that the isospin
dependence of the mean field has a weak, if any, influence on
isotropy ratios.
Experimental $R_E$'s for a number of HICs between various xenon
and tin isotopes corroborate this result; cf.\ Table I of
Ref.\ \cite{lehaut}.

The function $f$ is expanded onto a moving basis of coherent
states taken as normalized Gaussians ${\cal G}_\chi$ (${\cal G}_\phi$)
with frozen width $\chi$ ($\phi$) in $\bf r$ ($\bf p$) space:
\begin{equation}
f({\bf r},{\bf p};t) = \frac{A}{N} \sum_i 
  {\cal G}_\chi({\bf r}-{\bf r}_i) {\cal G}_\phi({\bf p}-{\bf p}_i) .
\label{df}
\end{equation}
\noindent
$A$ is the system mass number and $N$ is the number of
coherent states ($N/A$ equals 60 in the present study).
The widths $\chi$ and $\phi$ are chosen such as to best
reproduce the nuclear ground state characteristics of the
two colliding nuclei.
The local density reads
\begin{equation}
\rho({\bf r}) = \int {\rm d}^3{\bf p} ~ f({\bf r},{\bf p}) .
\label{ro}
\end{equation}
\noindent
Gaussians move in the self-consistent mean field and
suffer hard scattering between them, controlled by the
Uehling-Uhlenbeck collision integral accounting for the
fermionic character of interacting particles\
\cite{ueh33}:
\begin{eqnarray}
I_{\rm{coll}} & = & \frac{4g}{m^{2}}
       \int {\rm d}^3{\bf p_2}~{\rm d}^3{\bf p_3}~{\rm d}^3{\bf p_4}~
       \frac{d\sigma^m_{\rm NN}}{d\Omega}~
       \nonumber\\
& \times & \delta({\bf p + p_2 - p_3 - p_4})~
       \delta(\epsilon+\epsilon_2-\epsilon_3-\epsilon_4) \\
& \times & \lbrack (1-\bar f)(1-\bar {f_2}){f_3}{f_4}-
       (1-\bar {f_3})(1-\bar {f_4}){f_2} f \rbrack ,
       \nonumber
\label{CIeq}
\end{eqnarray}
\noindent
which takes into account energy and momentum conservation
as well as the Pauli exclusion principle.
Here, $m$ denotes the nucleonic mass,
$\bar f=(2\pi \hbar)^{3} f({\bf r},{\bf p};t)/g$ is the
occupation number with $g$ the spin-isospin degeneracy\
\cite{seb89}, ${\bf p}$ and ${\bf p_2}$ (${\bf p_3}$ and
${\bf p_4}$) are initial (final) momenta of the scattering
particle pair, $\epsilon = \epsilon (p)$ is the
single-particle energy, while $\sigma^m_{\rm NN}$ is the
\textit{in-medium} nucleonic cross section.
$\sigma^m_{\rm NN}$ is scaled so that a Gaussian-averaged
mean-free path is the same as for a nucleon.
The cross section dependence on isospin has been reported as
crucial for the study of stopping\ \cite{liu01}.
Thhis kind of $\sigma_{\rm NN}$ parametrization was proposed
by Chen \textit{et al.}\ \cite{chen68}, which hereafter we
label $\sigma^{\rm Chen}_{\rm NN}$.
This phenomenological $\sigma^{\rm free}_{\rm NN}$ is based
on the empirical isospin and energy dependence of the free
$NN$ scattering and has been used in both\ \cite{lopez14}
and\ \cite{liu01}.
$\sigma_{\rm NN}$ due to Li and Machleidt\ \protect\cite{li93-4},
which accounts for the in-medium effects and is also isospin
dependent, is tested too.

\section{\label{sec3}Results and discussion}
Our stopping observable, the energy-based isotropy ratio, is
defined as the ratio between transverse $E_{\rm tran}$ and
longitudinal $E_{\rm long}$ energy components of reaction
ejectiles\ \cite{lehaut,lopez14}
\begin{equation}
R_E = \frac{\sum E_{\rm tran}^{i}}{2\,\sum E_{\rm long}^{i}} ,
\label{r_e}
\end{equation}
\noindent
where summation runs over particles of those reaction events that
satisfy certain selection criteria.
For the LV simulation results, the summation index $i$ of
Eq.\ (\ref{r_e}) runs over the free Gaussians, i.e.,
those which are not bound in large (residue-like) fragment(s).
Among the experimentally studied systems, the
$^{129}$Xe\,+\,$^{120}$Sn reaction has been measured at by
far the most abundant number of $E_{\rm inc}$ values\
\cite{lehaut,lopez14}.
Consequently, in the present work only the simulation of this
system is performed.
To acquire stable $R_E^{\rm sim}$ values, the simulation is
carried out up to 600 fm/$c$ at the lowest
$E_{\rm inc}$\,=\,12$A$\,--\,32$A$ MeV and up to 240 fm/$c$
at the highest $E_{\rm inc}$\,=\,80$A$\,--\,100$A$ MeV.
Beyond that time the calculation was continued until 8\,000
fm/$c$, considering only the Coulomb repulsion due to reaction
residues.
Special care was taken in order to perform our analysis
of simulation data as close as possible to experimental
conditions.

\subsection{\label{sec31}A density-dependent $NN$ cross section}
In the experimental analysis\ \cite{lopez14}, event selection
is based on the charged particle's multiplicity.
The authors selected the most central events that are estimated, in
cross-section units, to be equivalent to 50 to 150 mb\
\cite{lopez14}.
We adopt the median value of 100 mb for our analysis.
This amount corresponds to about 2\,\% of the total reaction
cross section $\sigma_R$ and, in a geometrical sharp-cut
approximation, to $b\!\leq$\,2.0 fm.
Consequently, in this subsection our simulation is limited to
$b\!\leq$\,2.0 fm.

\begin{figure}[h]
    \noindent
    \begin{minipage}[t]{\textwidth}
    \noindent
    \parbox{102mm}{
\includegraphics[width=102mm]{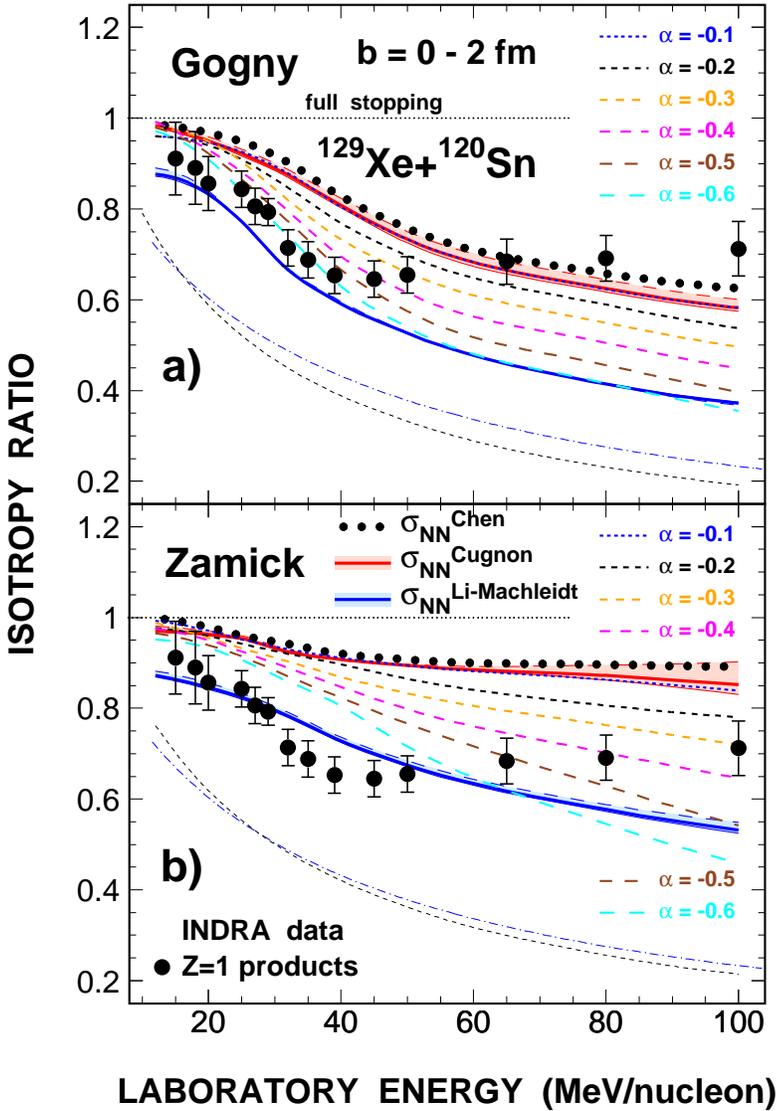}
    }
    \hfill
    \noindent
    \parbox{68mm}{
\caption{(Color online.)
Landau-Vlasov simulation results on the dependence of the
transverse-to-longitudinal energy ratio
$R_E^{\rm sim}$ of Eq.\ (\protect\ref{r_e}) as a function of
incident energy for the central $^{129}$Xe\,+\,$^{120}$Sn
reaction and several parametrizations of $\sigma^m_{\rm NN}$.
Panel (a) displays the results with the momentum dependent
D1-G1 EOS and panel (b) those with the zero-range Zamick EOS.
Heavy dotted curves represent the $R_E^{\rm sim}$ excitation
function obtained with the $\sigma^{\rm free}_{\rm NN}$ due
to Chen \textit{et al.}\ \protect\cite{chen68}.
The gray zones show the range of the
$R_E^{\rm sim}$-values limited by the impact parameter
$b$\,=\,1 fm (dashed bordering curves) and $b$\,=\,2 fm (full
bordering curves), while the heavy curve in each zone
represents the $b$-weighted $R_E^{\rm sim}$-values in the
range $b$\,=\,0\,--\,2 fm for $\sigma^m_{\rm NN}$ due to
Cugnon \textit{et al.}\ \protect\cite{cug87} (red curves and zones)
and Li and Machleidt\ \protect\cite{li93-4} (blue curves and
zones).
Thin dashed curves are obtained with the parametrization
of $\sigma^m_{\rm NN}$ due to Klakow \textit{et al.}\
\protect\cite{klakow} where the results for the different
values of the parameter $\alpha$ of Eq.\ (\protect\ref{kwb})
are distinguished by the varied dash size and color.
The filled circles and associated errors stand for the
$Z$\,=\,1 experimental $R_E^{\rm exp}$ values\
\protect\cite{lopez14}.
The thin dotted horizontal line denotes the full stopping value.
The entrance channel values of $R_E$ are shown by i) the thin
dashed curve resulting from the two Fermi spheres
($E_F$\,=\,38 MeV) displaced
for the entrance channel relative momentum and ii) by the thin
dash-dotted curves for the LV model values at the contact of
colliding nuclei for each of the two EOS used.
\label{snn}}
    }
    \end{minipage}
\end{figure}

Figure \ref{snn} displays $R_E^{\rm sim}$ obtained with the
momentum-dependent D1-G1 EOS (upper panel) and with the
zero-range Zamick EOS (lower panel) for several
parametrizations of the in-medium corrected $\sigma_{\rm NN}$.
For comparison, the experimental $R_E^{\rm exp}$'s are shown
by filled circles with the corresponding errors\ \cite{lopez14}.
As a reference, the $R_E^{\rm sim}$ results obtained with the
in-medium \textit{uncorrected} empirical free scattering
$\sigma^{\rm free}_{\rm NN}\!=\!\sigma^{\rm Chen}_{\rm NN}$\
\cite{chen68} are displayed by the heavy dotted curves.
This empirical $\sigma_{\rm NN}$ is used as an input for the
in-medium modified $\sigma^m_{\rm NN}$ suggested by Cugnon
\textit{et al.}\ \cite{cug87}.
In their Brueckner $G$-matrix in-medium renormalization of
the $NN$ interaction, they obtained a set of parameters
explicitly describing the dependence of
$\sigma^{\rm Cugnon}_{\rm NN}$ on the local density\
\cite{cug87}.
These simulation results are displayed by the red curves and
reddish zone in Fig.\ \ref{snn}: the zone shows the range of
the $R_E^{\rm sim}$ values limited by the impact parameters
$b$\,=\,1 fm (dashed bordering curve) and $b$\,=\,2 fm (full
bordering curve).
($b$\,=\,0 fm has no weight and $R_E^{\rm sim}$ at most of
energies is roughly the same for $b$\,=\,0 and 1 fm.)
The heavy curve in each zone represents the $b$-weighted
$R_E^{\rm sim}$ value in the range $b$\,=\,0\,--\,2 fm and
corresponds to 2\,\% of $\sigma_R$.
For both EOS, the $R_E^{\rm sim}$ values with
$\sigma^{\rm Cugnon}_{\rm NN}$ are very similar to those
obtained with $\sigma^{\rm Chen}_{\rm NN}$ (dotted curves).
Clearly, in the full energy range investigated here the
in-medium effects of $\sigma^{\rm Cugnon}_{\rm NN}$ have
rather weak impact on the $R_E$ observable.
Consequently, as for $\sigma^{\rm Chen}_{\rm NN}$, the
compatibility of $R_E^{\rm sim}$ and $R_E^{\rm exp}$ for
both EOS may be observed at the lowest $E_{\rm inc}$ only
when the experimental errors are accounted for.
In addition, for the Zamick EOS, Fig.\ \ref{snn}b), the
simulation strongly overshoots the data at the highest
$E_{\rm inc}$'s.

A full \textit{ab initio} microscopic study of
$\sigma^m_{\rm NN}$ based upon the Dirac-Brueckner approach
to nuclear matter was performed by Li and Machleidt\
\cite{li93-4}.
Besides dependence on energy, isospin, and density of
$\sigma^{\rm Cugnon}_{\rm NN}$ for this
$\sigma^{\rm Li-Machleidt}_{\rm NN}$, we have added an
explicit dependence on angle. 
In contrast to the scattering of neutrons, which is taken as
isotropic, those between neutron and proton $\sigma_{\rm np}$
and between protons $\sigma_{\rm pp}$ are anisotropic in
accordance with the fit of Ref.\ \cite{seb07}, which is given
in detail in the Appendix.
Similarly to above, the corresponding values of $R_E^{\rm sim}$
are displayed by the blue curves and bluish zone in Fig.\
\ref{snn}.
Again, the compatibility of $R_E^{\rm sim}$ with $R_E^{\rm exp}$
is unsatisfactory.
Nevertheless, for the D1-G1 EOS and
$E_{\rm inc}\!\lesssim\!E_F$, Fig.\ \ref{snn}(a), the slope
of the isotropy ratio excitation function is correct but
the simulation somewhat undershoots the experimental points:
$R_E^{\rm sim}$ may be taken as compatible with the lower
edges of experimental errors on $R_E^{\rm exp}$.
For the Zamick EOS, Fig.\ \ref{snn}(b), the compatibility
with $R_E^{\rm exp}$ exists at low $E_{\rm inc}$ and around
$E_{\rm inc}\!\sim$\,60$A$ MeV, but the general features of
the data are poorly reproduced.
Manifestly, none of the above parametrizations of
$\sigma^m_{\rm NN}$ and EOS can account for the observed
behavior of the $R_E$ stopping observable in the full energy
range.

The parameters in the above $\sigma^m_{\rm NN}$ are of a
fixed value.
By an expansion around the saturation value $\rho_0$, Klakow
\textit{et al.}\ have suggested a simple parametrization for the
dependence of $\sigma^m_{\rm NN}$ on the evolving nuclear
density\ \cite{klakow},
\begin{equation}
\sigma^m_{\rm NN} = \sigma^{\rm free}_{\rm NN}
        ( 1 + \alpha \,\frac{\rho}{\rho_0} ) ,
\label{kwb}
\end{equation}
\noindent
where $\rho$ is evaluated locally according to Eq.\ (\ref{ro}),
and $\alpha$ is a free parameter assumed to reduce the cross
section, thus it is strictly negative.
As before, for $\sigma^{\rm free}_{\rm NN}$ the value taken
is the empirical $\sigma^{\rm Chen}_{\rm NN}$.
The authors have recommended for $\alpha$ the domain
[--0.3, --0.1]\ \cite{klakow}.
In our simulation $\alpha$ is varied between --\nolinebreak0.1 and --0.6.
These $R_E^{\rm sim}$ are presented in Fig.\ \ref{snn} by
the thinner dashed curves with variable dash size.
They display a more or less regular dependence on both
$E_{\rm inc}$ and $\alpha$.
For the nonlocal EOS and --\nolinebreak0.6\,$\le$\,$\alpha$\,$\le$\,--0.5 the
$R_E^{\rm exp}$ values at $E_{\rm inc}\!\lesssim\!E_F$ are
well reproduced in both slope and absolute value; cf.\
Fig.\ \ref{snn}(a).
At energies higher than $E_F$, however, for each $E_{\rm inc}$
another and regularly increasing value of the parameter
$\alpha$ is required such that, at the highest $E_{\rm inc}$
here considered, it should become positive, implying an
in-medium enhancement rather than a reduction of
$\sigma^{\rm free}_{\rm NN}$ at
$E_{\rm inc}$\,$\gtrsim$\,80$A$ MeV.
Let us mention that $R_E^{\rm sim}$ with
$\sigma^{\rm Cugnon}_{\rm NN}$ corresponds to that of
$\sigma^{\rm Klakow}_{\rm NN}$ with
$\alpha$\,=\,--0.1 in the full range of $E_{\rm inc}$
considered and for both EOS.
Simulation results with $\sigma^{\rm Li-Machleidt}_{\rm NN}$
and D1-G1 EOS are compatible to $\sigma^{\rm Klakow}_{\rm NN}$
with $\alpha$\,=\,--0.6 and $E_{\rm inc}$\,$\gtrsim$\,50$A$ MeV.
For the Zamick EOS of Fig.\ \ref{snn}(b) one does not find a
range of $E_{\rm inc}$ of stable value of the parameter
$\alpha$ that gives

    \onecolumngrid
    \twocolumngrid

\noindent
$R_E^{\rm sim}$'s compatible with either
$R_E^{\rm exp}$ or those $R_E^{\rm sim}$ due to
$\sigma^{\rm Li-Machleidt}_{\rm NN}$.

In conclusion, neither choice of $\sigma^m_{\rm NN}$
allows for a unique description of experimental observation.
One faces the fact that every model study, ours and previous\
\cite{su13,liu01,zhang,kaur11,vin12,zhao}, fails to reproduce
with a single set of parameters the INDRA experimental
results in the full energy range studied\ \cite{lehaut,lopez14}.
In particular, all models but\ \cite{zhao} predict steadily
decreasing values of $R_E^{\rm sim}$ when $E_{\rm inc}$ increases,
while experimental $R_E^{\rm exp}$ results display a break in
the slope around the Fermi energy $E_F$.

\subsection{\label{sec32}Global modification of the free $NN$ cross section}
Being clearly unable to reproduce the experimental data with
different parametrizations of the residual $NN$ cross section,
with or without momentum dependence of the force,
let us concentrate on our second task that is, by following
Ref.\ \cite{lopez14}, to infer the multiplicative factor
${\cal F}$ between the in-medium $NN$ cross section
$\sigma^m_{\rm NN}$ and the free $\sigma^{\rm free}_{\rm NN}$ one:
\begin{equation}
\sigma^m_{\rm NN} = {\cal F} \,\sigma^{\rm free}_{\rm NN} .
\label{sig_m}
\end{equation}
\noindent
As previously done and as in\ \cite{lopez14}, we take
$\sigma^{\rm free}_{\rm NN}\!=\!\sigma^{\rm Chen}_{\rm NN}$\
\cite{chen68}.
Of course, this simple cross-section normalization factor
${\cal F}$ cannot completely describe the rather complex modification
of the free $NN$ interaction occurring in the nuclear medium.
In particular, such a $\sigma^m_{\rm NN}$ is frozen during a
reaction course and depends only indirectly on $E_{\rm inc}$.
Nevertheless, the prescription of Eq.\ (\ref{sig_m}) allows
one to get an insight into the global in-medium effects on
nuclear medium stopping properties and enables a comparison of
the factor ${\cal F}_{\rm sim}$ obtained in our simulation with
${\cal F}_{\rm exp}$ of Ref.\ \cite{lopez14}.

\begin{figure}[tb]
\includegraphics[width=\columnwidth]{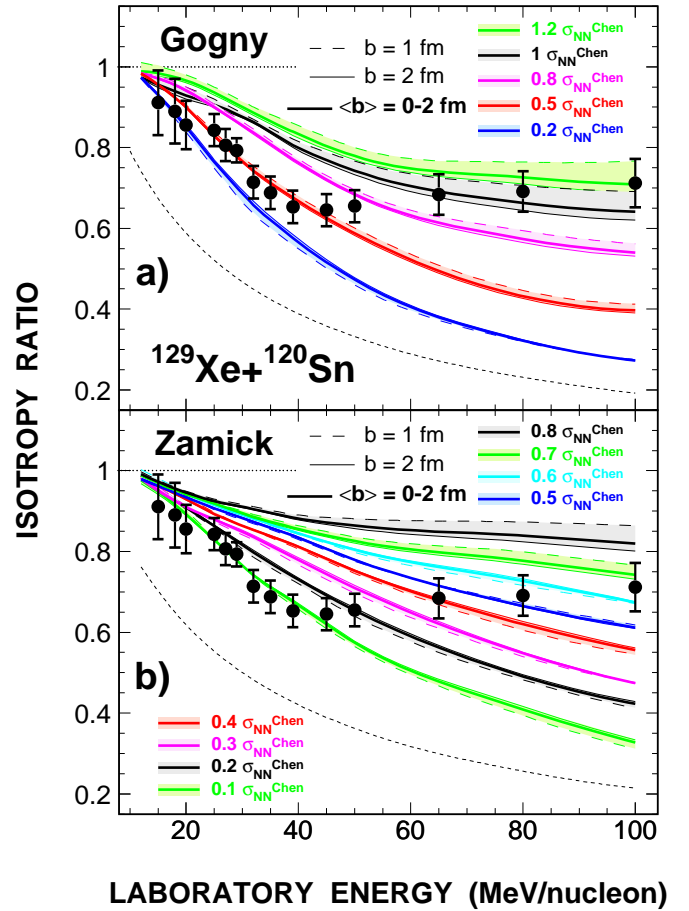}
\caption{(Color online.)
$R_E^{\rm sim}$ as a function of incident energy for the
central Xe\,+\,Sn reaction and several
values of the $\sigma^{\rm free}_{\rm NN}$ scaling factor
${\cal F}$ of Eq.\ (\protect\ref{sig_m}).
Upper (lower) panel shows results obtained with the D1-G1
(Zamick) EOS.
The colored zones and curves have the same
meaning as in Fig.\ \protect\ref{snn}, but here for the
scaled $\sigma_{\rm NN}$ of Chen \textit{et al.}\ \protect\cite{chen68}.
For more details see the caption of Fig.\ \protect\ref{snn}
and the text.
\label{sf}}
\end{figure}

Figure \ref{sf} displays $R_E^{\rm sim}$ as a function of
$E_{\rm inc}$ and the $NN$ cross-section factor ${\cal F}$ for
the two effective interactions.
In the D1-G1 EOS case, Fig.\ \ref{sf}(a), the parameter
${\cal F}$ takes values 0.2, 0.5, 0.8, 1.0, 1.2, and 1.5.
For the Zamick EOS, Fig.\ \ref{sf}(b), it is varied between
0.1 and 0.8 in steps of 0.1.\footnote{%
 Simulation was also performed for the pure mean field,
 i.e., the Vlasov equation with zeroed right-hand-side of
 Eq.\ (\protect\ref{LVeq}), which is equivalent to taking
 the parameter ${\cal F}$\,=\,0.}
As in Fig.\ \ref{snn}, $R_E^{\rm sim}$ are for central HIC
with $b$\,$\leq$\,2.0 fm, where $b$\,=\,1 fm (2 fm) results
are represented by the thin dashed (full) curves that boarder
the (colored) zone of each of the ${\cal F}$ values.
As before, the heavy curve in each zone shows the
$b$-weighted $R_E^{\rm sim}$ that corresponds to 2\,\% of
$\sigma_R$.
$R_E^{\rm sim}$ displays a regular dependence on $E_{\rm inc}$
and ${\cal F}$.
In accordance with expectation and corroborating the results
of Fig.\ \ref{snn}, higher $\sigma^m_{\rm NN}$ (larger
${\cal F}$) implies higher stopping power of HICs.
Unlike experimental $R_E^{\rm exp}$ and like our results of
Fig.\ \ref{snn}, as well as of a number of previous theoretical
works\ \cite{su13,liu01,zhang,kaur11,vin12,zhao}, the LV-simulation
$R_E^{\rm sim}$ steadily decreases with $E_{\rm inc}$ for all
${\cal F}$ without a minimum around $E_F$.
At the lowest $E_{\rm inc}$'s the mean field completely
dominates the course of the collision, and for each
${\cal F}$ value and both EOS $R_E^{\rm sim}$ is compatible
with $R_E^{\rm exp}$.
For $E_{\rm inc}\!\leq$\,45$A$ MeV, $R_E^{\rm exp}$ is
well reproduced by the ${\cal F}$\,=\,0.5 curve [Fig.\
\ref{sf}(a)] and by the ${\cal F}$\,=\,0.1 one [Fig.\
\ref{sf}(b)], respectively.
Again, a single value of ${\cal F}$ cannot reproduce
experimental results.
However, similarly to the case of the parameter $\alpha$ of
Eq.\ (\ref{kwb}), by allowing ${\cal F}$ to change with
$E_{\rm inc}$ one may find a set of ${\cal F}$ values to
achieve an agreement between $R_E^{\rm sim}$ and
$R_E^{\rm exp}$.
The behavior of both the parameter $\alpha$ and the factor
${\cal F}$ with $E_{\rm inc}$ corroborates the experimental
finding\ \cite{lopez14} that the effective in-medium cross section
$\sigma^m_{\rm NN}$ drastically changes with $E_{\rm inc}$
and that around $E_F$ there is a break in this dependence.

\begin{figure}[tb]
\includegraphics[width=\columnwidth]{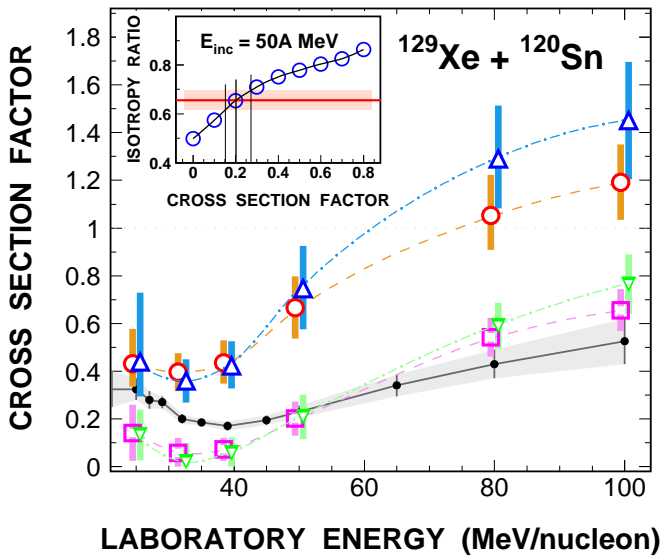}
\caption{(Color online.)
Cross section correction factor ${\cal F}$ of Eq.\
(\protect\ref{sig_m}) as a function of $E_{\rm inc}$.
Open circles and squares interpolated by the
dashed curve display ${\cal F}_{\rm sim}$ values obtained
for $R_E^{\rm sim}$ evaluated in the $b$\,=\,0\,--\,2 fm
range using the D1-G1 EOS and Zamick EOS, respectively.
Open triangles, upright and reversed, interpolated by the
dash-dotted curve denote
the Gaussian weighted $R_E^{\rm sim}$ evaluated in the
interval $b$\,=\,0\,--\,5 fm
for the D1-G1 EOS and Zamick EOS, respectively.
The symbols (but not the curves) are slightly
shifted in $E_{\rm inc}$ to avoid overlapping error bars.
Full dots, full curve, and gray zone represent
${\cal F}_{\rm exp}$ values and their uncertainty deduced
from the experimental $R_E^{\rm exp}$\ \protect\cite{lopez14}.
All curves are merely intended to guide the eye.
The inset explains the procedure used to extract the values
of ${\cal F}_{\rm sim}$.
For more details see text.
\label{cf}}
\end{figure}

We take the $b$-weighted $R_E^{\rm sim}$ as the starting
point to infer information about the correction factor
${\cal F}$ by which one would have to multiply
$\sigma^{\rm Chen}_{\rm NN}$ to comply with $R_E^{\rm exp}$.
The procedure is evidenced in the inset of Fig.\ \ref{cf}
in which the D1-G1 EOS at 50$A$ MeV is shown as an example.
The horizontal red line and reddish background zone display
the $R_E^{\rm exp}$ value and its uncertainty, respectively,
at 50$A$ MeV.
Blue circles joined by a broken line are the LV simulation
$R_E^{\rm sim}$ as a function of ${\cal F}$ at the same energy.
The crossing of this broken line with the red line and the
edges of the reddish zone give the most appropriate value
for the factor ${\cal F}$ of Eq.\ (\ref{sig_m}) and its
uncertainty, respectively.

In the main panel of Fig.\ \ref{cf} we show, by the open circles and
squares joined by dashed curves, the thus obtained ${\cal F}$
values plotted against $E_{\rm inc}$ for the D1-G1 and
Zamick EOS, respectively.
Within experimental errors, the $R_E^{\rm sim}$ values for
$E_{\rm inc}\!\leq$\,20$A$ MeV are roughly compatible with
any ${\cal F}^{\rm exp}$ value and are not reported.
The LV model with the highly recommended nuclear interaction
D1-G1 for the range of energies of the present study and with
the empirical $NN$ cross section $\sigma^{\rm Chen}_{\rm NN}$
predicts, for all energies studied, about twice higher
${\cal F}$ values compared to those suggested by Fig.\ 10 of
Ref.\ \cite{lopez14}; these are presented in Fig.\ \ref{cf}
as black filled circles, with the gray area showing their
uncertainties.
In contrast to this, when experimental and simulation
uncertainties are accounted for, the zero-range
(\textit{local}) Skyrme interaction in the Zamick
implementation ${\cal F}_{\rm sim}$ is compatible with
${\cal F}_{\rm exp}$ above $E_{\rm inc}\!\approx$\,35$A$ MeV.
Let us underline that ${\cal F}_{\rm sim}$ for both EOS
display a minimum around $E_F$.
The minimum is relatively more pronounced than the one
suggested by ${\cal F}_{\rm exp}$ and it is somewhat shifted
in energy.
The Zamick EOS gives a ${\cal F}_{\rm sim}$ that reduces the
free $\sigma_{\rm NN}$ at all $E_{\rm inc}$ while the D1-G1
EOS gives ${\cal F}_{\rm sim}\!>$\,1 at
$E_{\rm inc}\!\gtrsim$\,80$A$ MeV.

\subsection{\label{sec33}Centrality versus multiplicity}
The most evident difference between a simulation and an
experimental data analysis is in the reliability of the
assessments of reaction impact parameter $b$.
Experimental selection of the most central collisions is made
by assuming that there is a biunivocal correspondence between
the reaction violence, i.e., the multiplicity of particles in
a reaction event, and the reaction centrality.
In a simulation the centrality is an input variable, thus it
is under full control.
In comparing simulation results and the earlier INDRA study
of $R_E$ and $R_p$\ \cite{lehaut} it has been underlined that
selecting events via multiplicity strongly mixes events of
different impact parameters over a rather broad span in $b$\
\cite{zhang,bonnet}.
Thus, let us examine the $b$ vs multiplicity relationship
and its influence on the isotropy ratio.
For that purpose we use the semidynamical general-purpose 
event-generating code HIPSE (Heavy-Ion Phase-Space Exploration)
intended to describe HICs at intermediate energies\
\cite{hipse}.
At each energy 100\,000 events are generated in the range
$b$\,=\,0\,--\,7 fm.
Let us note that, according to the expression of $\sigma_R$ in
Ref.\ \cite{tripathi}, the above range in $b$ is equivalent
to 0.27\,$\sigma_R$\,--\,0.30\,$\sigma_R$, depending on
$E_{\rm inc}$.
At $E_{\rm inc}$\,=\,50$A$ MeV the simulation was performed
in the full impact parameter range of the
$^{129}$Xe\,+\,$^{120}$Sn reaction, i.e., $b$\,=\,0\,--\,13
fm, in order to verify that in the non-covered range
($b$\,=\,7\,--\,13 fm) the high multiplicity events, in which
we are interested, are not present.
By passing the generated events through a sophisticated
INDRA-device geometry and detection-acceptance filter\
\cite{paolo} we found that it has no appreciable effect on
the $Z$\,=\,1 $R_E$ values.
Mostly, the change in $R_E$ due to this filter is below 0.5\,\%.

\begin{figure}[t]
\includegraphics[width=78mm]{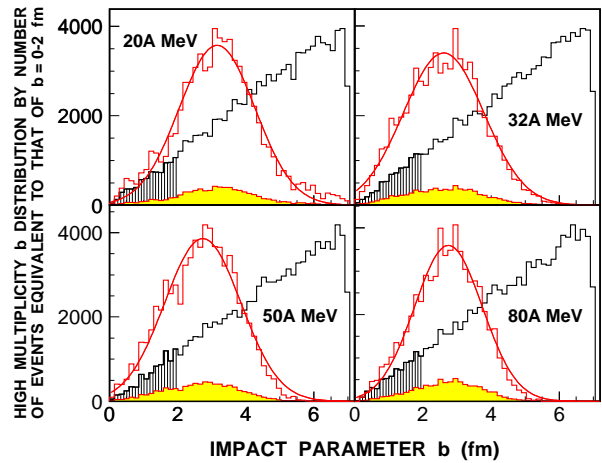}
\caption{(Color online.)
Simulation of the $^{129}$Xe\,+\,$^{120}$Sn reaction with the
HIPSE code\ \protect\cite{hipse}.
Shown are the $b$ distributions of the full data set (black line) and of
the high multiplicity subset (filled and hollow red-line
histograms), which is by number of events equivalent to the
one of $b$\,=\,0\,--\,2 fm (hatched part of the black
histograms) for four $E_{\rm inc}$.
The curves are the best fits by a Gaussian function.
For more details see text.
\label{eq_m}}
\end{figure}

Selecting the range $b$\,=\,0\,--\,2 fm, in a
geometrical sharp-cut approximation, corresponds to
1.82\,\% to 1.92\,\% of $\sigma_R$ in the studied
$E_{\rm inc}$ range, i.e., between 104 and 116 mb.
These values fall in the middle of the cross section values
of the selected subset of the most violent INDRA data
events analyzed in Ref.\ \cite{lopez14}.
Ideally, the reactions with $b\leq$\,2 fm should
correspond to 8163 out of the total 100\,000 generated events.
In reality, there were on the average 8109 such events with
a fluctuation up to 3\,\% from energy to energy.
We denote this precise number of events $N_{\rm 0-2}$ to
search for, in the full ensemble of 100\,000
events, the subset of events with the highest multiplicity
that is by number of events closest to $N_{\rm 0-2}$.
By $M_{\rm 0-2}$ we label both the lowest multiplicity
of the thus selected subset as well as the subset of events
itself at each $E_{\rm inc}$.

Let us check the behavior of the most violent $M_{\rm 0-2}$
HIPSE events.
As a kind of "background", in Fig.\ \ref{eq_m} we show
by the thin black line the $b$-distribution histogram of the
full 100\,000 event data set for each second studied $E_{\rm inc}$.
The $b$ distribution of the $M_{\rm 0-2}$ events is shown by
the red-line yellow-filled histogram.
These high-multiplicity events are generated in a large
domain of $b$ values which extends up to 5 fm.
To make the $M_{\rm 0-2}$ $b$ distribution better visible,
it is enlarged to the full frame size by the hollow red-line
histogram.
A Gaussian fit to it clearly demonstrates that the
normal-law of data statistics correctly reproduces the
distribution of $M_{\rm 0-2}$ subset over $b$'s.
These events are in minor part (3\,\% to 29\,\%) belonging
to the $b\!\leq$\,2 fm subset of the full data set (hatched
part of black histogram).
From the Gaussian fit one infers that the maximum of these
high-multiplicity events is about $b\!\approx$\,3 fm and
that it slightly decreases with the increasing $E_{\rm inc}$.

Finally, let us apply the HIPSE $M_{\rm 0-2}$ $b$ distribution
to the LV simulation results.
Taking the Gaussian fit values of Fig.\ \ref{eq_m} as the
weights for the integer values of $b$, the $b$-averaged
$R_E^{\rm sim}$ are obtained for each studied value of the
factor ${\cal F}$ of Eq.\ (\ref{sig_m}).
By this method, for the ${\cal F}$\,=\,1 case these
$R_E^{\rm sim}$ are, in millibarn units, also equivalent to
0.02\,$\sigma_R$.
${\cal F}^{\rm sim}$ extracted from thus averaged
$R_E^{\rm sim}$ is in Fig.\ \ref{cf} shown by
dot-dashed curves and open triangles, upright and reversed,
for the D1-G1 EOS and Zamick EOS, respectively.
For the nonlocal D1-G1 EOS the two $b$-averaging intervals
give strictly the same $R_E^{\rm sim}$ for
$E_{\rm inc}\!\leq$\,50$A$ MeV.
At $E_{\rm inc}$\,=\,80$A$ and 100$A$ MeV the respective
$R_E^{\rm sim}$ values differ by about 20\,\% but are
mutually compatible when errors are accounted for.
For the zero-range Zamick EOS in the full $E_{\rm inc}$
interval, two $b$-averaging intervals give compatible
predictions for the $\cal F_{\rm sim}$ values although for
$E_{\rm inc}\!\gtrsim\!E_F$ the more stringent centrality
results are in somewhat better agreement with the
${\cal F}_{\rm exp}$ values.

\section{\label{sec4}Summary and conclusions}
The semiclassical Landau-Vlasov (LV) transport model was used to
study the energy-based isotropy ratio $R_E$ of Eq.\ (\ref{r_e})
for the $^{129}$Xe\,+\,$^{120}$Sn reaction in the wide incident
energy range 12\,$A\!\leq\!E_{\rm inc}\!\leq$\,100$A$ MeV.
The focus of the present work is twofold:

\noindent
(1) the search for the set of model ingredients which most
favorably describes the experimental values $R_E^{\rm exp}$
for the $Z$\,=\,1 species of Ref.\ \cite{lopez14} and

\noindent
(2) the comparison of the simulation multiplicative factor
${\cal F}$ representing the global in-medium change of the free 
$NN$ cross section $\sigma^{\rm Chen}_{\rm NN}$ of Ref.\
\cite{chen68} with the one deduced from the experimental
$R_E^{\rm exp}$\ \cite{lopez14}.

In approaching the above point 1 we investigated
(i) the role of the dependence of the nuclear mean field on
momentum, i.e., of the nonlocality of the interaction, and
(ii) the impact of the residual $NN$ interaction through
varied parametrizations of $\sigma_{\rm NN}$.
The success in reproducing the experimental isotropy ratios
$R_E^{\rm exp}$ of Ref.\ \cite{lopez14} is mixed:
Below the Fermi energy $E_F$, the LV model with the strongly
in-medium reduced $NN$ cross section $\sigma^m_{\rm NN}$ of
Refs.\ \cite{li93-4,klakow} and with the momentum-dependent
D1-G1 EOS leads to a correct description of $R_E^{\rm exp}$;
cf.\ Fig.\ \ref{snn}(a).
A similar result may be obtained with both the D1-G1 EOS and
the zero-range Zamick EOS when the free $NN$ cross section
is strongly scaled down by a constant multiplicative factor
${\cal F}$ of Eq.\ (\ref{sig_m}): for the D1-G1 EOS
${\cal F}\!\sim$\,0.5, Fig.\ \ref{sf}(a), and for the Zamick
EOS ${\cal F}\!\sim$\,0.1, Fig.\ \ref{sf}(b).
Above $E_F$ there is no unique set of model parameters
which would lead to a favorable description of the
experimental $R_E^{\rm exp}$.
Earlier studies of the observable $R_E$\
\cite{su13,liu01,zhang,kaur11,vin12,zhao} have also failed
to reproduce the experimental results of Ref.\ \cite{lehaut}.
We emphasize, however, that a smooth variation of
the parameter that controls the in-medium value of
$\sigma_{\rm NN}$ would lead to a complete description of the
experimental $R_E^{\rm exp}$ for both EOS.
Should one draw a conclusion that none of the existing studies
on the in-medium modifications of $\sigma_{\rm NN}$ around the
Fermi energy appropriately accounts for the physical reality?
The local or the nonlocal character of the interaction does
not elucidate this question.

Regarding the above point 2, we may summarize the outcome
of our study as follows:
the LV simulation with the local Zamick EOS predicts the
$NN$ cross-section correction factor ${\cal F}_{\rm sim}$
which clearly supports the experimentally deduced
${\cal F}_{\rm exp}$\ \cite{lopez14}.
The model predicts

\noindent
(1) an appreciable reduction of $NN$ cross section all along
the energy range of interest, as well as

\noindent
(2) the appearance of a break in the slope of the
multiplicative factor ${\cal F}$ after a minimum located near
$E_F$.

\noindent
However, the agreement or disagreement between the absolute
values of ${\cal F}_{\rm sim}$ and ${\cal F}_{\rm exp}$
should be considered with some caution due to two possible causes.
On one hand, the value of factor ${\cal F}$ may be altered by
reaction centrality.
Accordingly, an investigation of $R_E$ with a quasidynamical
event generator HIPSE\ \cite{hipse} was carried out.
It reveals that the event selection based on multiplicity
and the geometrical sharp-cut approximation is not a correct
centrality selector.
Indeed, corroborating earlier findings\ \cite{zhang,bonnet},
we show that this selection approach strongly mixes events of
different impact parameters over a rather broad span of $b$
values; cf.\ Fig.\ \ref{eq_m}.
When a properly weighted contribution of $b$'s involved in
the high-multiplicity events is accounted for, the isotropy
ratios calculated for the thus relaxed centrality requirement and
those strictly central do not differ much.
The thus extracted ${\cal F}_{\rm sim}$ does not change much
as well.
On the other hand, the derived ${\cal F}_{\rm exp}$ values
are based on a number of strong assumptions that allowed the
link between the stopping ratio $R_E$ and the in-medium $NN$
cross section\ \cite{lopez14}.
Hence, besides further experimental and theoretical
considerations of the stopping observable $R_E$ intended to
disentangle the remaining ambiguities a study of other
related observables may shed some fresh light on the subject.
In addition, the experimentally observed strong and rapid
change of the effective in-medium residual $NN$ cross section
beyond the Fermi energy urges for an \textit{ab initio}
theoretical analysis of this problem, the solution of which
might lie in the way the exclusion principle is accounted
for\ \cite{su13} and/or by incorporating the recent observation
of short-range correlations in nuclei\ \cite{subedi,hen}.
Their consequences for transport descriptions of heavy-ion
reactions are of high interest and need to be investigated.

\vspace*{-1.5ex}
\begin{acknowledgments}
Z.B.\ gratefully acknowledges the financial support and the
warm hospitality of the Facult\'e des Sciences of University
of Nantes and the Laboratory SUBATECH, UMR 6457.
This work has been supported in part by Croatian Science
Foundation under Project No. 7194 and in part by the
Scientific Center of Excellence for Advanced Materials and
Sensors.
\end{acknowledgments}

\setlength{\tabcolsep}{4.3mm}
\onecolumngrid
\begin{table*}
\caption{Coefficients of neutron-proton scattering angular
distribution function $f_{\rm np}$ of Eq.\ (\protect\ref{fnp})
as parametrized by Eqs.\ (\protect\ref{fnp_par}).
\label{tabnp} }
\begin{center}
\begin{tabular}{cccccccccc}
\hline
$E$ (MeV)           & $e_i$ (MeV) & Ref. & $a_{i,1}$ & $a_{i,2}$ & $a_{i,3}$ & $b_{i,1}$ & $b_{i,2}$ & $b_{i,3}$ & $c_i$ \\
\hline
$E\!<$\,26          &    0  &            & 0        & 0        & 1        & 0        & 0        & 0        &  1   \\
26\,$\le\!E\!<$\,35 &   26  & \cite{mon} & 0        & 0        & 1        &   0.966  &  -0.426  &   1.372  &  9   \\
35\,$\le\!E\!<$\,45 &   35  & \cite{ben} &   0.97   &  -0.426  &   2.372  &   0.32   &   0.35   &  -0.40   & 10   \\
45\,$\le\!E\!<$\,53 &   45  & \cite{ben} &   1.29   &  -0.073  &   1.97   &   0.32   &  -0.127  &  -0.18   &  8   \\
53\,$\le\!E\!<$\,63 &   53  & \cite{ben} &   1.61   &  -0.2    &   1.79   &  -0.04   &   0.51   &   0.16   & 10   \\
63\,$\le\!E\!<$\,73 &   63  & \cite{ben,sca,kin} &   1.57   &   0.31   &   1.95   &   0.33   &  -0.59   &  -0.30   & 10   \\
73\,$\le\!E\!<$\,90 &   73  & \cite{ben} &   1.9    &  -0.28   &   1.65   &   0.9    &   0.205  &  -0.16   & 17   \\
90\,$\le\!E\!<$\,130 &  90  & \cite{sca,chi} &   2.8    &  -0.075  &   1.49   &   1.2    &  -0.465  &  -0.094  & 40   \\
130\,$\le\!E\!<$\,319 & 129 & \cite{mea} &   4.0    &  -0.54   &   1.396  &  -1.3    &   0.665  &  -0.81   & 189  \\
$E\!\ge$\,319       &   319 & \cite{kee} &   2.69   &   0.125  &   0.588  & 0        & 0        & 0        &  1   \\
\hline
\end{tabular}
\end{center}
\end{table*}
\vspace*{-3.5ex}
\twocolumngrid
\appendix*
\section{}
The angular dependence of the nucleon-nucleon ($NN$) cross
section $\sigma_{\rm NN}$ is expressed as
\begin{equation}
\sigma(\theta)_{\rm c.m.} {\rm (mb/sr)} =
               f(E,\theta) \times
               \frac{\sigma_{\rm tot} {\rm (mb)}}{4\pi} ,
\end{equation}
\noindent
where $\sigma_{\rm tot}\!=\!\sigma^{\rm Li-Machleidt}_{\rm NN}$
is the total elastic cross section due to Li and Machleidt\
\protect\cite{li93-4}.
The dimensionless weighting factor $f(E,\theta)$
is equal to unity for the scattering between neutrons
($f_{\rm nn}\!\equiv$\,1) and is
increasingly anisotropic as energy increases for neutron-proton
scattering (the $f_{\rm np}$ case), and especially becomes
strongly forward-backward peaked for the scattering between
protons ($f_{\rm pp}$).
The parametrization of the angular dependence of $\sigma_{\rm np}$
is defined as\ \cite{seb07}
\begin{equation}
f(E,\theta)_{\rm np} = \frac{A_1 \cos^4 \theta - A_2 \cos^3 \theta + A_3}
                                     {A_1/5 + A_3} ,
\label{fnp}
\end{equation}
\noindent
where, for the purpose of the fitting, the coefficients $A_k$ at each
energy are expressed by the following functional dependence:

\setlength{\tabcolsep}{1.8mm}
\begin{table}[hb]
\caption{Coefficients of proton-proton scattering angular
distribution function $f_{\rm pp}$ of Eq.\ (\protect\ref{fpp})
as parametrized by Eqs.\ (\protect\ref{fpp_par}).
\label{tabpp} }
\begin{center}
\begin{tabular}{cccccc}
\hline
$E$ (MeV)         & $\epsilon_i$ (MeV) & Ref. & $\alpha_{i,1}$ & $\alpha_{i,2}$ & $\alpha_{i,3}$\\
\hline
$E\!<$\,5           &         0  &            &     5176.1     &      -8.91     &  100.0  \\
5\,$\le\!E\!<$\,9.9 &         5  & \cite{spp} &     5176.1     &      -8.91     &  100.0  \\
9.9\,$\le\!E\!<$\,19.7 &     9.9 & \cite{ja1} &     1795.6     &      -9.29     &   52.62 \\
19.7\,$\le\!E\!<$\,39.4 &   19.7 & \cite{ja2} &     1071.0     &      -12.0     &   24.95 \\
39.4\,$\le\!E\!<$\,68 &     39.4 & \cite{joh} &     1382.2     &      -19.26    &   11.16 \\
68\,$\le\!E\!<$\,144  &      68  & \cite{you} &     1880.5     &      -26.77    &    6.16 \\
$E\!\ge$\,144         &     144  & \cite{ja3} &     4008.8     &      -45.92    &    3.99 \\
\hline
\end{tabular}
\end{center}
\end{table}

\begin{figure}[h]
\includegraphics[width=78mm]{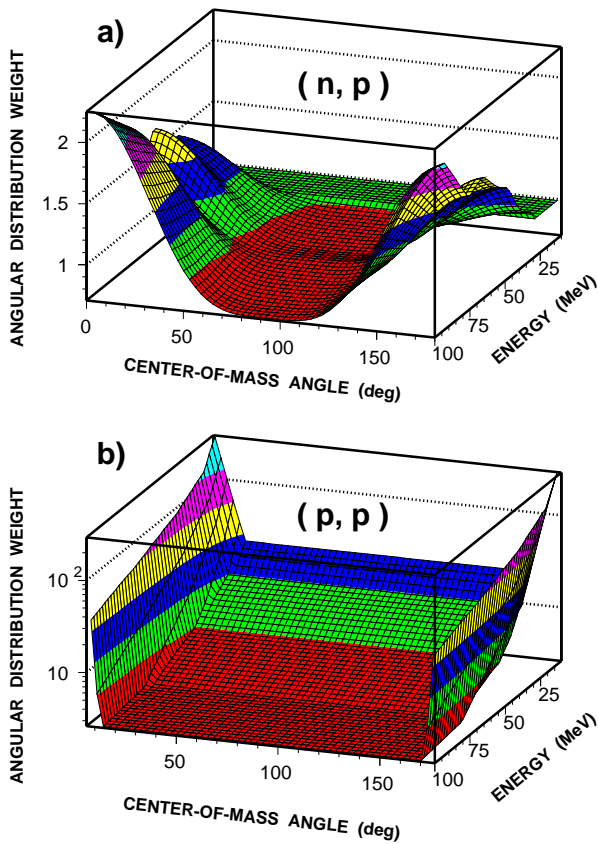}
\caption{(Color online.)
Dimensionless weighting factor $f(E,\theta)$
which modulates total elastic cross section as a function
of polar angle $\theta_{\rm c.m.}$ and nucleon incident
energy $E$ for the scattering of neutron and proton (upper
panel) and between protons (lower panel), where the
applicate axis is in logaritmic scale.
\label{ad}}
\end{figure}

\begin{equation}
A_k(E) = a_{i,k} + \frac{b_{i,k} (E-e_i)}{c_i}, ~~~ k = 1,2,3 .
\label{fnp_par}
\end{equation}
\noindent
Parameters $a_{i,k}$, $b_{i,k}$, and $c_i$ are fixed by
fitting the experimental $\sigma_{\rm np}$ data at nine
beam energies $e_i$ between 26 and 319 MeV, index $i$
running over energies.
$E$ and $e_i$ are expressed in MeV units.
Between these $e_i$ values the parameters are assumed to
change linearly with $E$.
The values of these parameters are given in Table\ \ref{tabnp}
and $f(E,\theta)_{\rm np}$ is shown in Fig.\ \ref{ad}(a).

Adopting a very crude estimate, the polar angle dependence
of $\sigma_{\rm pp}$ is defined as\ \cite{seb07}
\[
f(E,\theta)_{\rm pp} = \left\{ \begin{array}{ll}
            B_1 \exp (B_2 \theta),       & \theta<\theta_0, \\
            B_3,                         & \theta_0\!<\!\theta\!<\!\pi-\theta_0, \quad\quad\quad\\
            B_1 \exp (B_2 (\pi-\theta)), & \theta>\pi-\theta_0, \\
                               \end{array} \right.
\]\\[-14.3ex]
\begin{equation}
\label{fpp}
\end{equation}

\vspace*{3.5ex}
\noindent
where coefficients $B_k$ are expressed by the following
functional dependence:
\begin{equation}
B_k(E) = \alpha_{i,k} + \frac{(\alpha_{i+1,k}-\alpha_{i,k}) (E-\epsilon_i)}
                             {\epsilon_{i+1}-\epsilon_i}, ~~~ k = 1,2,3 .
\label{fpp_par}
\end{equation}
\noindent
Due to indistinguishability of particles, coefficients $B_1$ and
$B_3$ are divided by 2.
At each energy $E$ the limiting angle reads
$\theta_0\!=\!\ln(B_3/B_1) / B_2$.
The overall angular distribution normalization is given by
the value of $\sigma^{\rm Li-Machleidt}_{\rm pp}$, and Eq.\
(\protect\ref{fpp}) is used to define, on Monte Carlo
grounds, the angle into which a couple of charged Gaussians
is scattered in a $p-p$ collision.
The parameters $\alpha_{i,k}$ are fixed by fitting
experimental differential cross sections $\sigma_{\rm pp}$
at six energies $\epsilon_i$ ranging from 5 to 144 MeV
denoted by the index $i$.
As above, $E$ and $\epsilon_i$ are expressed in MeV units.
Between these energies, parameters are assumed to change
linearly with $E$.
The values of these parameters are given in Table\ \ref{tabpp}
and $f(E,\theta)_{\rm pp}$ is shown in Fig.\ \ref{ad}(b).

\end{document}